\documentclass[aip,rsi,reprint,graphicx]{revtex4-1}

\draft 

\usepackage{amssymb}
\usepackage{xcolor}
\usepackage{graphicx}
\usepackage{float}
\usepackage{booktabs} 
\usepackage{subfig}

\usepackage{dcolumn}
\usepackage{bm}
\usepackage[utf8]{inputenc}
\usepackage[T1]{fontenc}
\usepackage{mathptmx}
\usepackage{etoolbox}
\usepackage{attachfile}
\usepackage[normalem]{ulem}

\definecolor{NAL_blue}{HTML}{004C97}

\newcommand{\jpl}[1]{{#1}}

\newcommand{\rff}[1]{{#1}}
\newcommand{\yw}[1]{{#1}}
\renewcommand\sout[1]{}

\newcommand{\ywsecond}[1]{{#1}}
\newcommand{\jplsecond}[1]{{#1}}
\newcommand{\rffsecond}[1]{{#1}}
\newcommand{\soutsecond}[1]{}

\begin{document}


\title{Low latency optical-based mode tracking with machine learning deployed on FPGAs on a tokamak} 



\author{Y. Wei}
\email[]{yw2714@columbia.edu}
\affiliation{Department of Applied Physics and Applied Mathematics, Columbia University, New York, New York 10027, USA}

\author{R. F. Forelli}
\email[]{rforelli@fnal.gov}
\affiliation{Real-time Processing Systems Division, Fermi National Accelerator Laboratory, Batavia, Illinois 60510, USA}
\affiliation{Department of Electrical and Computer Engineering, Lehigh University, Bethlehem, Pennsylvania 18015, USA}

\author{C. Hansen}
\affiliation{Department of Applied Physics and Applied Mathematics, Columbia University, New York, New York 10027, USA}

\author{J. P. Levesque}
\affiliation{Department of Applied Physics and Applied Mathematics, Columbia University, New York, New York 10027, USA}

\author{N. Tran}
\affiliation{Real-time Processing Systems Division, Fermi National Accelerator Laboratory, Batavia, Illinois 60510, USA}
\affiliation{Department of Electrical and Computer Engineering, Northwestern University, Evanston, Illinois 60208, USA}

\author{J. C. Agar}
\affiliation{Department of Mechanical Engineering and Mechanics, Drexel University, Philadelphia, Pennsylvania 19104, USA}

\author{G. Di Guglielmo}
\affiliation{Microelectronics Division, Fermi National Accelerator Laboratory, Batavia, Illinois 60510, USA}
\affiliation{Department of Electrical and Computer Engineering, Northwestern University, Evanston, Illinois 60208, USA}

\author{M. E. Mauel}
\affiliation{Department of Applied Physics and Applied Mathematics, Columbia University, New York, New York 10027, USA}

\author{G. A. Navratil}
\affiliation{Department of Applied Physics and Applied Mathematics, Columbia University, New York, New York 10027, USA}


\date{\today}


\begin{abstract}

Active feedback control in magnetic confinement fusion devices is desirable to mitigate plasma instabilities and enable robust operation.  Optical high-speed cameras provide a powerful, non-invasive diagnostic and can be suitable for these applications.  In this study, we process \yw{\sout{fast}high-speed} camera data, at rates exceeding 100~kfps, on \textit{in situ} field-programmable gate array (FPGA) hardware to track magnetohydrodynamic (MHD) mode evolution and generate control signals in real-time.  
Our system utilizes a convolutional neural network (CNN) model which predicts the $n$=1 MHD mode amplitude and phase using camera images with better accuracy than other tested non-deep-learning-based methods. By implementing this model directly within the standard FPGA readout hardware of the high-speed camera diagnostic, our mode tracking system achieves a total trigger-to-output latency of 17.6~\mbox{$\mu$s} and a throughput of up to 120~kfps.
This study at the High Beta Tokamak-Extended Pulse (HBT-EP) experiment demonstrates an FPGA-based high-speed camera data acquisition and processing system, enabling application in real-time machine-learning-based tokamak diagnostic and control as well as potential applications in other scientific domains.

\end{abstract}

\pacs{}

\maketitle 



\section{Introduction}\label{intro}


High-speed optical cameras with intelligent, machine learning (ML)-enabled real-time processing and control capabilities bring entirely new capabilities to a wide-range of scientific applications from cell-sorting to microscopy to fusion~\cite{10.3389/fdata.2022.787421}.   We present an open-source workflow accessible to domain scientists that embeds a deep learning model in the data acquisition and control loop field-programmable gate arrays (FPGAs). We demonstrate this workflow for the application of microsecond-level low latency tracking of magnetohydrodynamic (MHD) instabilites in tokamak plasmas. This system \jplsecond{\soutsecond{will be used} is being prepared to use} for real-time instability control in upcoming experiments on the High Beta Tokamak-Extended Pulse (HBT-EP) device.


While magnetic confinement devices can be designed to operate far from stability limits with nominal plasma conditions, transient events or other deviations from desired parameters may lead to instability. Efficient usage of the applied magnetic field to confine higher plasma pressure generally leads to less stable equilibria due to $\beta$-driven instabilities (\yw{where beta is the ratio between the plasma and magnetic pressures, or } $\beta = \frac{\mathrm{nT}}{\mathrm{B}^2/2\mu_0}$\yw{, $n$ and $T$ are the plasma density and temperature, and $B$ is the magnetic field strength}). As tokamaks and other magnetic confinement devices achieve fusion-relevant conditions, real-time diagnosis and control of plasma instabilities is essential for maintaining desirable plasma parameters and robust performance. Plasma conditions and instabilities in tokamaks can evolve on an Alfv\'enic timescale for the most demanding cases, necessitating fast evaluations and control decisions on a timescale of microseconds for quickly growing or rotating kink and tearing instabilities \cite{chu_ppcf_2010}. For this reason, real-time control of plasma properties requires reasonably accurate implementation of control algorithms with latencies commensurate with the plasma dynamics.


Recently, deep learning \cite{LECUN15} and other data-driven machine learning techniques have been applied to a variety of areas in plasma  and fusion research. For tokamak control applications, current studies have mainly focused on the prediction and control of disruptive events and equilibrium quantities. 
These studies can be broadly divided into two categories: 1. by developing a data-driven model to predict the intended plasma quantities \cite{Jalalvand21IEEE, Boyer21NF, Zhu23NF, Harbeck19Nature} and using these predictions to inform a non-machine-learning-based control algorithm \cite{Abbate23JPP, Wei21NF}, or 2. by training a deep reinforcement learning agent on a computational \cite{Degrave22Nature, Dubbioso23FED} or data-driven \cite{Seo22NF, Char23PMLR} simulator and using the trained agent to directly generate control policies. 
Some of these trained models have also been tested in real-time and have achieved latencies on the order of the device's \yw{plasma control system} cycle time, ranging from milliseconds to as low as 50 $\mu$s \cite{Abbate23JPP, Boyer21NF, Shin22NF, Char23PMLR, Degrave22Nature}.

Besides disruption and equilibrium, prediction and control of non-equilibrium quantities such as MHD instabilities \cite{Piccione22NF, Wei23PPCF} and \ywsecond{\soutsecond{ELMs}edge localized modes} \cite{Shin22NF, Song23NET} using machine learning approaches have also been investigated. However, applying these models for real-time control can be more challenging as the model needs to perform inference with latency
similar to or smaller than the timescale of the\ywsecond{\soutsecond{plasma event} \jplsecond{phenomena} to be controlled}, which can be as fast as\jplsecond{\soutsecond{in the microsecond or nanosecond range} a few microseconds}. Depending on the complexity of the specific algorithm, a real-time implementation may require special optimization and computing hardware in order to achieve the target latency.

In this paper, we present the first real-time low-latency application of a deep learning algorithm for MHD mode control on a tokamak device. 
This work goes beyond previous use of an FPGA \yw{for MHD mode control using magnetic sensors on HBT-EP} \cite{Klein2005PoP, Hanson2009RSI}, and combines efficient processing of FPGAs with high-speed imaging camera diagnostic and convolutional neural networks (CNNs). The successful tracking of $n$=1 long wavelength MHD modes on the HBT-EP experiment \cite{Wei23PPCF} is further developed and implemented on a FPGA chip onboard the Euresys frame grabber card in the existing camera diagnostic system.

\rff{FPGAs consist of reprogrammable hardware blocks of limited capacity, comprising various dedicated logic for specific functions such as multiply-accumulators for efficient in-memory compute. Thus, these devices offer the advantage of being well-suited for high-speed input/output (I/O) and streaming machine learning inference, as they can be optimized specifically for the computational patterns of deep neural networks to achieve low-latency, low-power inference~\cite{Aarrestad:2021zos} within the resource constraints of the device.}
FPGA-based computing platforms have enabled neural network models to perform online single-sample inference with latencies in the 100~ns to 10~$\mu$s range~\cite{Aarrestad:2021zos}, far outperforming their counterpart GPU implementations. However, deploying a model on an FPGA device presents several challenges, including designing within the resource constraints of the FPGA chip. For this reason we utilized the High-Level Synthesis for Machine Learning (\texttt{hls4ml})~\cite{vloncar_2021_5680908} package to optimize our model before synthesizing and deploying the model firmware to the frame grabber's FPGA. Our implementation achieved a latency of 7.7 $\mu$s for the CNN model itself, 17.6 $\mu$s total for the acquisition trigger to output control request, and is capable of operating at pipelined frame rates of up to 120~kfps, satisfying the requirement for MHD mode control on HBT-EP. This is the first demonstration of 100\,kHz low-latency control algorithms for real-time MHD mode tracking using raw high-speed camera outputs in fusion experiments.

The layout of this paper is as follows.  In Section~\ref{sec:sysdesign}, we describe the HBT-EP\yw{\sout{Tokamak}} experiment and the\yw{\sout{fast} high-speed} camera system used for MHD mode tracking.  We detail the choice of readout technology used for implementing the control algorithm in real time and the tools we use for deploying machine learning algorithms in FPGA readout hardware.  In Section~\ref{sec:model}, we describe the deep learning algorithm optimization for the\yw{\sout{fast} high-speed} camera readout hardware and the performance of the algorithm. This includes trade-offs between computing resources, latency, and performance. Then, in Section~\ref{sec:implementation} we detail the final implementation and validation of the control algorithm deployed at the HBT-EP experiment.  Finally, we summarize and describe future work.

\section{System design}\label{sec:sysdesign}

\subsection{Diagnostics and hardware}

\begin{figure}
\includegraphics[width=0.99\linewidth]{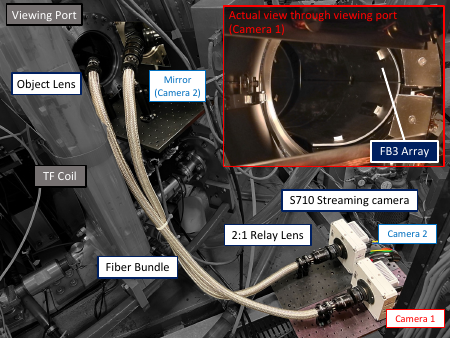}
\caption{Hardware setup of the high-speed camera diagnostics on HBT-EP as used during the previous run campaign \cite{Wei23PPCF}, with an actual view of the chamber (similar to Camera 1 view) shown in the top-right inset. 
\yw{\sout{The same database was used in this study for training and testing the various CNN models for FPGA implementation as described in the following sections.}}
Reused with permission from Y. Wei \textit{et al} (2023) \textit{PPCF} \textbf{65} 074002. Copyright 2023 IOP Publishing.}
\label{fig:diagnostic_setup}
\end{figure}

HBT-EP \cite{Maurer11PPCF} is a circular, ohmically heated, large aspect ratio tokamak (R/a $\sim$ 6). It has an on-axis toroidal magnetic field of 0.33~\mbox{T}, typical plasma currents of 15~\mbox{kA}, a major radius of 0.92~\mbox{m}, and a plasma minor radius of 0.15~\mbox{m}.  Figure~\ref{fig:diagnostic_setup} shows the HBT-EP high-speed camera diagnostic setup. 
The setup consists of two Phantom S710 cameras which are capable of operating at frame rates between 100-400~\mbox{kfps}. The cameras are placed away from the toroidal field (TF) coils and measure visible light emissions, predominantly of the D-alpha line from the plasma edge \cite{Angelini15PPCF}, through a midplane glass viewing port using a fiber bundle and relay lenses. This system offers high flexibility and has been used to study MHD activities during the “flat-top” period \cite{Angelini15PPCF} or during disruptions \cite{Saperstein23thesis} through the use of different camera views, frame rates, and exposure settings. 
\yw{The high-speed cameras have the advantage of being external optical diagnostics, allowing them to be easily adjusted and maintained independent of the machine’s maintenance cycles. For this camera-based real-time mode tracking and control system, achieving microsecond-level fast diagnostic response time requires only ex-vessel components and a single viewing window to observe the plasma, rather than any in-vessel components such as additional distributed in-vessel magnetic sensors.}

\begin{figure*}
\includegraphics[width=0.99\linewidth]{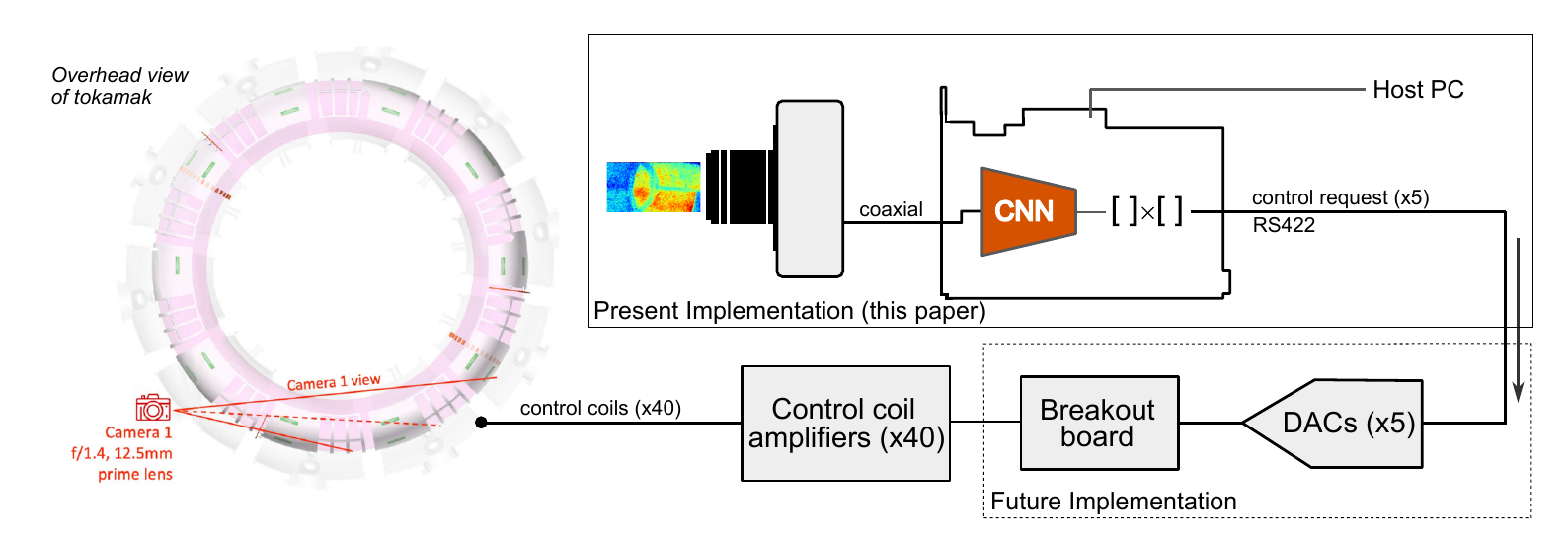}%
\caption{Data flow in the\sout{proposed} camera-and-FPGA-based active feedback control system. \jpl{The portion of the implementation detailed in this paper is within the solid box.  Control signals will be applied to plasmas in upcoming run campaigns.}
}
\label{fig:overview}
\end{figure*}

The Phantom S710 does not possess the capability to process the camera sensor readout;  instead, a dedicated PCIe frame grabber card is required to capture and interpret the raw camera data into monochrome images. We use a Euresys Coaxlink Octo frame grabber card which utilizes the CoaXPress camera interface standard and implements its image processing capabilities on a Xilinx Kintex UltraScale KU035 FPGA. Euresys enables the user to also implement custom firmware on the card's FPGA through a reference design toolkit called CustomLogic~\cite{customlogic_software}. For this project, we leveraged this toolkit to implement our CNN on the frame grabber's FPGA and subsequently write the computed control request over the frame grabber's RS422 I/O pins.  In placing the computation and output request generation on the acquisition card, we avoid the overhead associated with multiple PCIe hops, latency associated with standard processors, and the cost associated with developing a highly-custom solution.  Currently, CustomLogic only supports the use of a single frame grabber card and one of the two available CXP-6 connection banks on the Coaxlink Octo card. This restricts us to using only one of the two available cameras in this real-time system and limits the maximum achievable frame rate compared to the passive diagnostic setup.

Figure~\ref{fig:overview} shows the data flow of the \jpl{\sout{proposed}} camera-based control system on HBT-EP. We use a single camera (Camera 1) which captures a cross-sectional view of the plasma. The captured frames are streamed in real time to the frame grabber card, which processes the optical data, passes the images through the neural network model and generates new control requests. Currently, we measure these output control requests and other auxiliary signals \jpl{as digital pulses} on an oscilloscope to verify the timing and latency of the system. \jpl{The digital control requests can then be sent to digital-to-analog converters (DACs) which connect to amplifiers for generating the target mode shape on the radial field saddle coils~\cite{Maurer11PPCF}.  Closed-loop control experiments have not yet been conducted, though the design is ready for upcoming run campaigns.}


\subsection{Design considerations}

The \sout{proposed} real-time control system is subject to two major design constraints: the system latency and the available FPGA resources. 

Firstly, MHD modes on HBT-EP typically rotate at frequencies around 10 kHz which corresponds to a period of 100\mbox{ $\mu$s}. In order to suppress these modes, the control system needs to achieve a total input-to-out latency several times smaller than a rotation period and a sampling interval (i.e., the initiation interval, the frequency at which the system can process a new input) at a similar level, if not faster. Currently, the fastest implementation of the GPU-based control system on HBT-EP has achieved latencies around 16~\mbox{$\mu$s} and sampling intervals between 4--6~\mbox{$\mu$s} \cite{Rath13PPCF, Peng16PPCF}. This system, however, is significantly less computationally challenging compared to our \jpl{camera} setup as it takes in one or two orders of magnitude less input data with 40--80 magnetic sensor channels compared to 1,024--8,192 pixels (depending on the actual image resolutions) and uses a single matrix multiplication as opposed to a neural network to perform mode tracking.
These distinctions make achieving similar levels of latency and sampling intervals considerably more challenging for a high-speed camera and FPGA-based control system and have been our primary considerations during our design and optimization phases.

Secondly, the total resource capacity of our frame grabber card’s FPGA is relatively small compared to modern FPGA-based data center and accelerator cards used in high-throughput applications. This FPGA consists of just over 200,000 look-up tables (LUTs) and 1,700 digital signal processing (DSP) blocks, the two elements primarily used for matrix-vector multiplication arithmetic and logic, and 540 \yw{block random-access memories \sout{RAMs}} (BRAMs). \sout{In addition, approximately} Approximately 25\% of the LUTs, 10\% of the DSPs, and 35\% of the BRAM resources are dedicated to implementing the CoaXPress protocol and other proprietary frame grabber firmware. \rff{In addition, the manufacturer's reference design also defines a clock constraint of 250~MHz which becomes difficult to meet with larger designs relative to the chip area. These constraints} further \sout{restricts} \rff{restrict} our model size and potential for latency optimization.

Given these two major constraints, our goal in terms of the hardware co-design is to find a solution that achieves \ywsecond{\soutsecond{an optimal} a suitable} trade-off between the model’s prediction performance and its latency while fitting within the resource constraints of the frame grabber FPGA. The model \yw{development, }optimization and synthesis process using \texttt{hls4ml} is described in the following section.


\section{Neural network model co-design}
\label{sec:model}

\subsection{Training dataset}

The CNN models are developed using a dataset collected from a previous run campaign \cite{Wei23PPCF}. This dataset consists of camera frame images and the corresponding magnetic mode signals at 86,275 time slices (or \textit{samples}) over 45 discharges. These shots used a mode-control-type shot style on HBT-EP and contained both $m/n$ = 4/1 and 3/1 modes (where $m$ and $n$ are the poloidal and toroidal mode numbers respectively) rotating at around 8--12 kHz during different time periods of the the discharge. We divided the dataset the same way as in the previous study, by using the first 40 shots as the training set (76,975 samples) and the last 5 shots as the testing set (9,300 samples). The training set is further split randomly by samples using a 9:1 ratio into training and validation \ywsecond{\soutsecond{for hyperparameters tuning} sets for tuning the model hyperparameters}.


The frame images were captured by the two Phantom S710 cameras through an optical view port (Figure~\ref{fig:diagnostic_setup}). Both cameras had a cross-sectional view of the plasma toward opposite toroidal directions. The images were taken at 250~\mbox{kfps}, $128\times64$ pixels (width$\times$height) resolution, 12-bit grayscale bit depth, and 3~\mbox{$\mu$s} exposure time. For this study we only used the images from Camera 1 to train the CNN models due to the present constraint of CustomLogic, although it has been shown that using images from both cameras slightly improved the accuracy of the model's predictions \cite{Wei23PPCF}.


For the ground truths (or training labels), we used the sine and cosine components of the $n=1$ mode determined from an in-vessel low-field-side magnetic sensor array (FB3). \yw{These mode components were calculated using Fourier decomposition through least squares fitting.}\yw{\sout{This sensor array} The FB3 as well as other arrays} has been used extensively during previous mode control experiments and for evaluating other non-magnetic mode tracking methods, therefore by accurately reproducing these signals the camera-based mode tracking system should achieve similar mode suppression results. In future iterations of this system, the ground truth information may be acquired from alternative diagnostic or numerical source that requires extensive post-shot analysis in order to better leverage the strengths of CNNs for image regression and low-latency inference through FPGAs.



\subsection{Model development and optimization strategy}

\begin{figure*}
    \centering
    \includegraphics[width=0.70\linewidth]{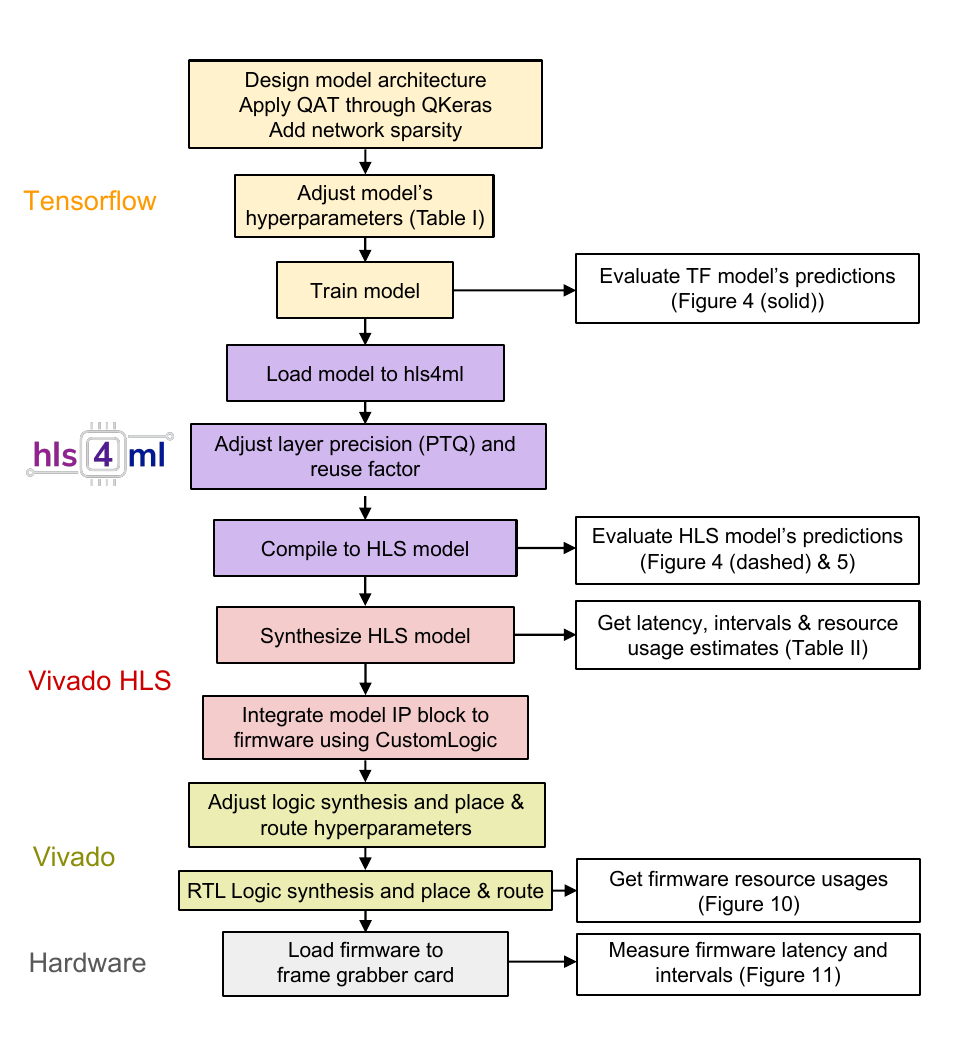}
    \caption{\yw{Illustration of the firmware design procedures.  Results of each step impact choices made during repeated iterations through the process.}}
    \label{fig:system-design}
\end{figure*}

\yw{Using this dataset, we developed the mode tracking CNN models in Tensorflow \cite{tensorflow2015-whitepaper}. Figure~\ref{fig:system-design} illustrates the overall workflow for designing the model and subsequently its firmware implementation. Our model takes in a single camera frame image and predicts the sine and cosine components of the $n$=1 MHD mode. A typical CNN model consists of two parts: a feature extractor utilizing consecutive blocks of Conv2D--MaxPooling2D layers to identify dominant features from the input image, and a regressor using Dense layers to map the identified features to the desired output values. During the training process, a model's predictions are compared to the provided ground truth values, and their differences are used to update the network's trainable parameters through backpropagation. CNN models are especially suitable for processing imaged-based data. From our previous study, we found that our implemented model was not only capable of tracking the evolution of the $n$=1 MHD mode on HBT-EP, but also outperformed other, more conventional, algorithms using identical image inputs \cite{Wei23PPCF}.}

\yw{In this study, we implemented the trained CNN models in the readout path of Camera 1 in the Euresys Coaxlink Octo frame grabber card. To do so, we utilized the \texttt{hls4ml} package to compile a Tensorflow model into a high-level synthesis (HLS) representation.}
\yw{\sout{We implemented the CNN model in the readout path of Camera 1 in the Euresys Coaxlink Octo frame grabber card, and utilized the \texttt{hls4ml} package to compile the model, originally implemented in Tensorflow \cite{tensorflow2015-whitepaper}, into a high-level synthesis (HLS) representation.}}
This compiled HLS model is then synthesized to a register-transfer level (RTL) description using the Xilinx Vivado Design Suite~\cite{xilinx} and subsequently deployed to the target FPGA device. 

\sout{To tailor our CNN model for the previously mentioned HBT-EP system constraints, \texttt{hls4ml} offers a variety of optimizations to meet hardware resource and latency requirements.} \rff{ We utilized the resource and latency optimization features offered by \texttt{hls4ml} to tailor our CNN model for the HBT-EP system constraints.} Efficient model design for hardware is a multi-dimensional design space from CNN hyperparameter choices to hardware microarchitecture implementation.  On the hardware implementation side, there are three primary high-level optimizations \rff{in \texttt{hls4ml} that we utilized here}\sout{available to the user in \texttt{hls4ml}}.  First, using full 32-bit floating point input data and parameter representation for the model in hardware is unnecessary.  Therefore, \texttt{hls4ml} utilizes arbitrary-precision fixed-point representations more suited for FPGA hardware to store model weights and activations. \rff{We took} \sout{The user can take} advantage of post-training quantization (PTQ) to appropriately select the weight and accumulator bit widths \textit{after} training the model.  More optimally, \rff{quantization can be accounted} \sout{the user can account} for \sout{quantization} \textit{during training} using quantization-aware training (QAT) frameworks like QKeras~\cite{qkeras}, Brevitas~\cite{brevitas}, and HAWQ~\cite{hawq}.  In this study, we \rff{used} \sout{use} the QKeras framework, built on Keras~\cite{chollet2015keras}, and TensorFlow for customizable quantization.  QAT often yields more optimally (lower precision) trained neural networks over PTQ. 

In addition to quantizing the parameters, \rff{we reduced} the model’s resource usage\sout{can be reduced} further by applying sparsity to its weights and biases. \rff{We achieved this}\sout{This can be done} using tools like the Pruning API in the TensorFlow Model Optimization Toolkit~\cite{tfmot} to arbitrarily force a certain percentage of the parameters to be zero during training. \texttt{hls4ml} \sout{can} \rff{accounts} \sout{account} for these zero parameters during its conversion process, thus reducing the number of parameters needed to implement each of the pruned layers. 

\rff{Finally, we leveraged hls4ml’s “reuse factor” tunable parameter (i.e. how many times a logic element is reused during one inference) to take advantage of the highly parallel architecture of the FPGA fabric.} \sout{Finally, \texttt{hls4ml} leverages the highly parallel architecture of the FPGA fabric through a ``reuse factor'' tunable parameter (i.e. how many times a logic element is reused during one inference).} This option \rff{granted}\sout{grants the user} \rff{us} direct control over the degree of parallelization of the multiply-accumulate operations on a per-layer basis, enabling fine control over latency-area optimization. Additionally, dataflow pipelining enables further parallelization at the functional layer level, significantly reducing overall latency to nearly match the highest consumer. We\sout{utilize} \rff{utilized} this set of features, along with our own HLS and hardware optimizations outlined below to implement our CNN under challenging area and latency constraints. 

\subsection{Selected candidate models}
\label{subsec:candidates}

Using these optimization techniques with the Tensorflow, QKeras, and \texttt{hls4ml} tool flows, we leverage the co-design process to select a candidate model for implementation. All of these models share a similar network architecture to the single-camera single-frame model as described in Ref. \citenum{Wei23PPCF}, in using three Conv2D-MaxPool layers followed by two hidden dense layers and the output layer. \yw{This architecture and the associated hyperparameters were identified in the previous study through manual tuning. In this study, we decided to maintain this network architecture and focus on searching for firmware-relevant hyperparameters. Through this work we arrived at the following three representative candidate models which are described below. \sout{We describe each of these representative candidate models below.}}The relevant metrics for these models are compared in Table~\ref{tab:modelcomparison}.


\small
\begin{table}
\centering
\begin{ruledtabular}
\begin{tabular}{lcccc} 
\toprule
\textbf{Model Name} & \textbf{PPCF23} & \textbf{QAT+Pruning} & \textbf{Optimized} \\
\hline
Image Resolution & $128\times64$ & $128\times64$ & $32\times32$ \\
Conv Layer Filters & \{8, 8, 16\} & \{8, 8, 16\} & \{16, 16, 24\} \\
Dense Layer Widths & \{256, 64\} & \{256, 64\} & \{42, 64\} \\
Total Parameters & 362,730 & 362,730 & 12,910 \\
Parameter Precision & PTQ, 18 bits & QAT, 8 bits & QAT, 7 bits \\
Sparsity & none & 80\% & 50\% \\
Bit Operations & 3.62e13 & 6.86e12 & 2.85e11 \\
Reuse Factors & \begin{tabular}{@{}c@{}}\{1, 1, 1, \\ \yw{96}, 8, 1\}\end{tabular} & \begin{tabular}{@{}c@{}}\{1, 4, 16, \\ 48, 64, 128\}\end{tabular} & \begin{tabular}{@{}c@{}}\{1, 4, 16, \\ 48, 64, 128\}\end{tabular} \\
\bottomrule
\end{tabular}
\end{ruledtabular}
\caption{Comparison of hyperparameters and configurations for the \textbf{\texttt{PPCF23}} (baseline), \textbf{\texttt{QAT+Pruning}}, and \textbf{\texttt{Optimized}} models described in Section \ref{subsec:candidates}. Reuse factors selected for each layer of the networks are shown in the order: conv0, conv1, conv2, dense0, dense1, dense2. A lower value indicates a higher level of parallelization. 
}
\label{tab:modelcomparison}
\end{table}
\normalsize

\begin{itemize}
    \item The baseline implementation is called \textbf{\texttt{PPCF23}} after Ref. \citenum{Wei23PPCF} and uses the training hyperparameters from the original single-camera single-frame model. We applied minimal modifications to this implementation by only scanning the model and layer bit widths through PTQ to maintain performance with respect to the original 32-bit floating point model. 
    
    \item \textbf{\texttt{QAT+Pruning}}: A modified implementation based on the original model. For this model we applied QAT by replacing the Conv2D, ReLU \yw{activation}, and hidden Dense layers with the corresponding quantized layer from the QKeras library \yw{(QConv2D, QActivation, QDense)}. We scanned the quantization levels and sparsity using the training and validation sets; after determining the optimal hyperparameters, we trained the model one more time using the entire training dataset to obtain the final model parameters. 
    
    \item \textbf{\texttt{Optimized}}: A compressed and optimized implementation. This model uses a smaller input resolution of $32\times32$ as opposed to $128\times64$ used in all previous models, as well as fewer neurons at the first fully connected layer. The number of filters in the three QConv2D layers were also increased from \{8, 8, 16\} to \{16, 16, 24\} respectively 
    which from our tests were shown to improve the model's accuracy, potentially due to increased data flow into the downstream dense layers.
    These measures significantly reduced the number of model parameters by more than 96\%, not accounting for sparsity.
\end{itemize}

The list above does not outline all tested neural network models, but rather represents the most important optimizations that were performed in this study during the design space exploration.  The other crucial optimization on the hardware side we performed was determining a reasonable level of parallelization for each layer of the network through hls4ml's reuse factor parameter.  Fine-tuning these reuse factors can reduce FPGA resource usage while maintaining a sufficiently low model latency.

\begin{figure*}
\centering
\includegraphics[width=0.85\linewidth]{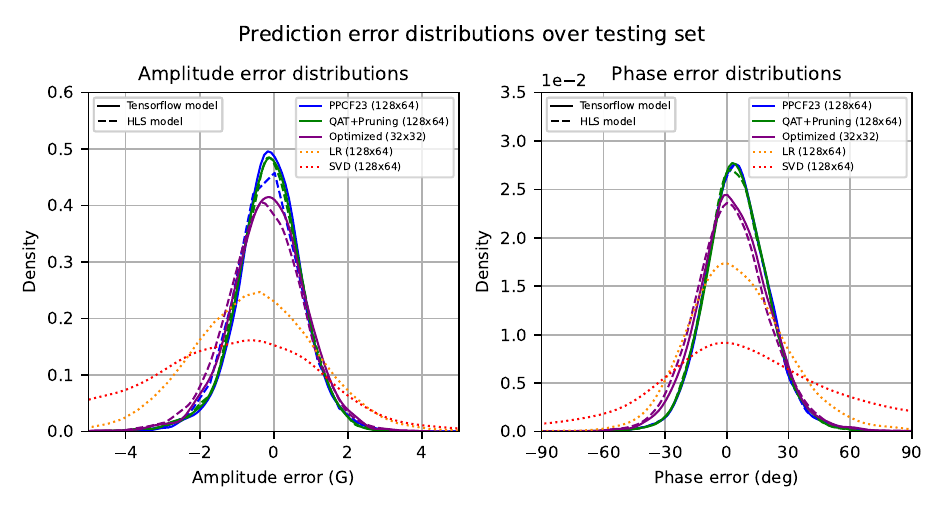}
\caption{\ywsecond{\soutsecond{Distributions} Probability distributions} of the amplitude (left) and phase (right) prediction errors over the testing set (ground truth amplitude~$>$3G), given by the three candidate CNN models. For each model the results of the Tensorflow implementation are shown in solid lines, and that of the converted HLS implementation are shown in dashed lines. Results of two additional non-deep-learning-based methods using linear regression and SVD-based algorithms \cite{Wei23PPCF} are included for comparison. All models except the \textbf{\texttt{Optimized}} model use input images of $128\times64$ pixels resolution, whereas the \textbf{\texttt{Optimized}} model uses $32\times32$ pixels resolution. \yw{Distributions are obtained using kernel density estimation.}
}
\label{fig:performance}
\end{figure*}

The first two implementations (\textbf{\texttt{PPCF23}} and \textbf{\texttt{QAT+Pruning}}) failed to achieve the targeted latency and resource criteria (see Table \ref{tab:finalmodel} and more discussion in Section \ref{subsec:model_results}). We have attempted to further adjust the hyperparameters of the \textbf{\texttt{QAT+Pruning}} model but did not arrive at a solution that satisfies our constraints.  
This was the primary reason why we decided to reduce the input dimensions in the \textbf{\texttt{Optimized}} implementation 
which significantly reduced the overall number of bit operations. 
\yw{To train this model, we resized input images from $128\times64$ to $32\times32$ pixels resolution in order to maintain the same field-of-view as the other two models. Note that this resizing step is not necessary for a future implementation which could instead directly capture in the desired resolution setting and changes the field-of-view through using an object lens with the appropriate focal length. For this reason we do not account the resource usages and latency of this resizing step in this model's firmware. }\ywsecond{Afterward, we adjusted the layer's hyperparameters as described in the table to obtain a solution that is realizable on our FPGA hardware with a reasonable performance. Through these adjustments, we arrived at the \texttt{Optimized} model which demonstrates a slightly reduced prediction accuracy but achieves significant reductions in its resource usages compared to the previous two full-size models.}


\subsection{Model performance and implementation results}
\label{subsec:model_results}

\yw{We evaluated the prediction accuracy of these three candidate implementations using the testing set data. This was done by comparing a trained model's predictions given the testing data with the associated ground truths from the dataset.} \rff{hls4ml is capable of simulating models at the HLS level through C-simulation in Vivado HLS. This process involves compiling the HLS design using a standard C++ compiler, such as g++, and then executing the program on a CPU to validate the algorithm's behavior. Vivado HLS defines arbitrary precision fixed-point data types in C++ to accurately replicate the algorithm's output in hardware.} Figure~\ref{fig:performance} compares the amplitude and phase error distributions of the predictions given by these three models; the results of the simpler, non-deep-learning, linear regression model and SVD-based models from the previous study \cite{Wei23PPCF} are also included for references. These distributions are obtained using data from the five test discharges during the mode active periods (ground truth amplitude $>$3G) in accordance with our previous analysis method. For each of the three CNN models, both the results of their original Tensorflow implementation (solid lines) and those of the converted \texttt{hls4ml} implementation (dashed lines) are shown in these figures.
We observed slight degradation in model accuracy after this conversion step especially for the amplitude predictions.

Among the three CNN models, the results of the \ywsecond{\soutsecond{second} \texttt{QAT+Pruning}} model (green lines) effectively matched those of the original model (solid blue lines) from the previous study even after the conversion (dashed green lines). This demonstrates the effects of applying QAT and sparsity on a neural network, as the now-compressed model maintains its prediction accuracy despite having 80\% of its network parameters set to zero while the remaining parameters using 24 fewer bits than those of the full Tensorflow model. 
\ywsecond{The \texttt{Optimized} model (purple lines) performs slightly worse than the two full-sized models; however, its performance still yields a large improvement over the two non-neural-network-based algorithms.}

\begin{figure*}
\centering
\includegraphics[width=0.8\linewidth]{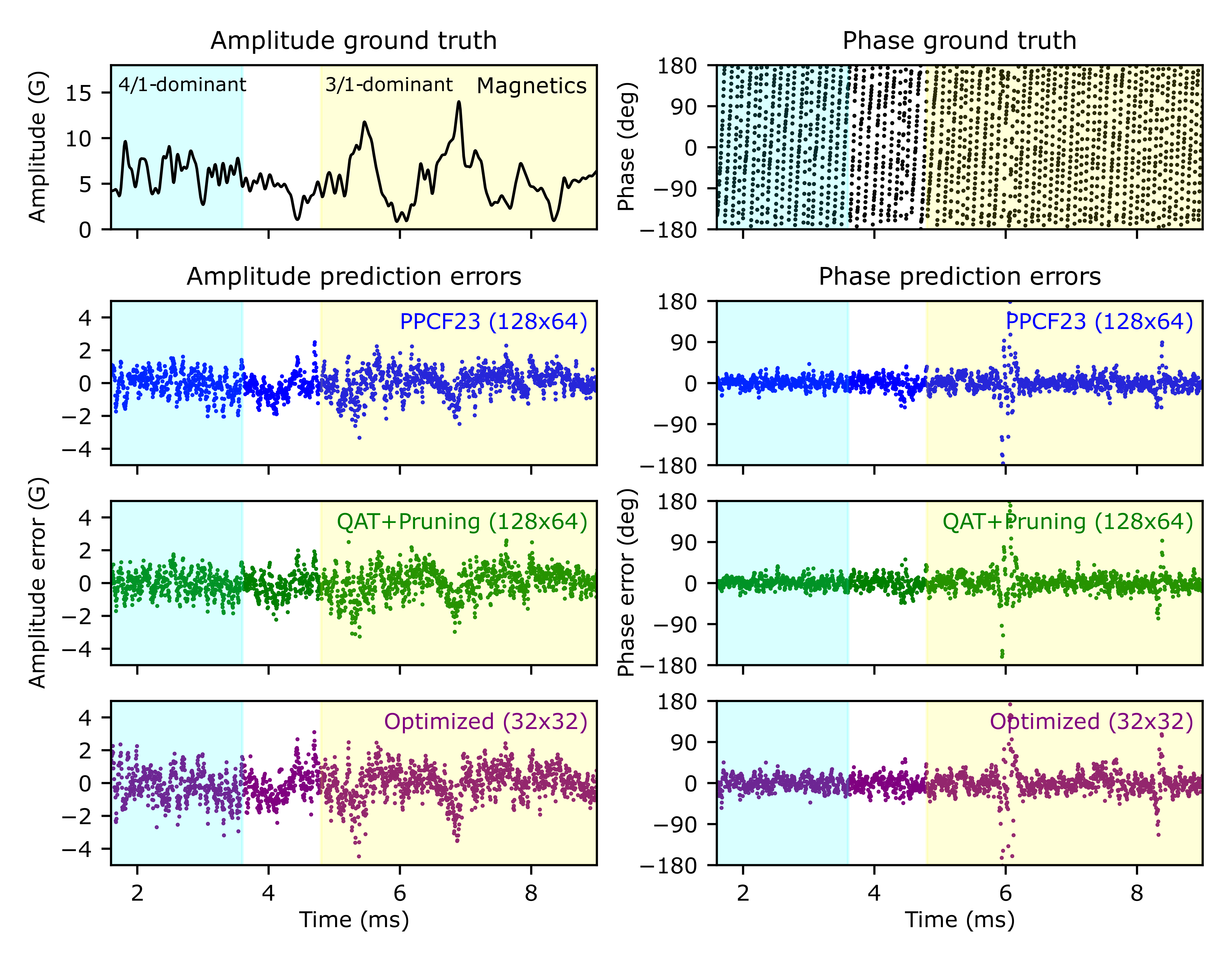}
\caption{Comparison of single shot tracking performance of the three candidate models after \texttt{hls4ml} conversion, using testing shot 114467. The 4/1 and 3/1-dominant periods are highlighted in cyan and yellow respectively. Top: Amplitude (left) and phase (right) ground truth calculated from magnetic sensors. Bottom 3 rows: Amplitude (left) and phase (right) prediction errors of the candidate models.
}
\label{fig:tracking}
\end{figure*}

A comparison of single-shot tracking performances using the three models after \texttt{hls4ml} conversion is shown in Figure~\ref{fig:tracking}. All three models are able to track both the amplitude and phase evolution across the entire discharge with reasonable accuracy. The phase predictions are only inaccurate when the mode amplitude was very small, such as at 6 and 8.4 ms in this shot. A slightly higher level of fluctuation is observed for the \textbf{\texttt{Optimized}} model in both its amplitude and phase predictions compared to the other two models. At this moment we do not believe this would affect the actual field applied to the plasma as these high frequency fluctuations would be ``smoothed out" by the control coil amplifiers and therefore should not affect the outcome in a control experiment; the exact effect will be investigated in the following study. 



\begin{figure}
\centering
\includegraphics[width=0.88\linewidth]{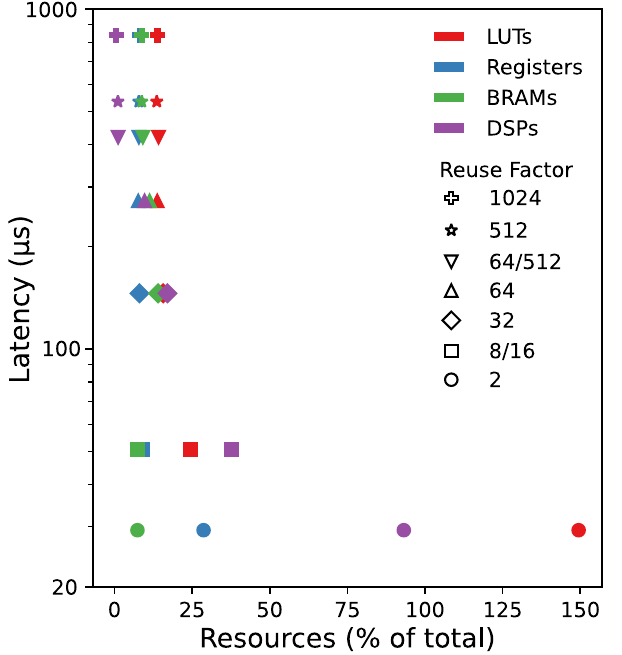}
\caption{Pareto frontier of optimized model's resource-latency configurations with increasing levels of parallelization. Colors represent the types of resources available on an FPGA: LUTs, Registers, BRAMs, and DSPs. Shapes represent different reuse factors. Resource usages are shown as \% of available resources on the KU035 FPGA. The final optimized model uses different reuse factors for different layers (see Table~\ref{tab:modelcomparison}).
}
\label{fig:rf}
\end{figure}

Using the \textbf{\texttt{Optimized}} model, we further optimized the hardware implementation by tuning the reuse factor of the model. Figure~\ref{fig:rf} shows the latency vs. FPGA resource usages of implementations using different reuse factors. We observe a trade-off between resource usages and computational latency as we increase the reuse factor. Through this approach, we select the proper reuse factor for each of the neural network layers to fit within the resource limits while keeping the latency as low as possible.


The estimated latencies, initiation intervals, and resource usages for this \textbf{\texttt{Optimized}} model and for the implementations of the two other models are shown in Table~\ref{tab:finalmodel}. \rff{Initiation interval denotes the amount of time the model must wait before accepting a new input.}
These estimates reflect only neural network logic usage and exclude the camera protocol and other logic. \rff{Estimated latencies and initiation intervals for all models in Table~\ref{tab:finalmodel} were obtained from RTL simulation and thus these designs did not have to fit within the chip resource constraints in order to obtain the benchmarks presented here.}
The implementation of the \textbf{\texttt{Optimized}} model achieved both our latency and resource targets.  In particular, the estimated latency of 6.6 $\mu$s is significantly lower than our target total latency of 16 $\mu$s, leaving a large margin for other processes such as image exposure, readout, and control request generation for integration with the control loop. The procedure for integrating this model with the frame grabber firmware is described in the next section.

\begin{table*}
\centering
\begin{ruledtabular}
\begin{tabular}{lcccccc} 
\toprule
 Model & Latency ($\mu s$) & Interval ($\mu s$) & BRAM & DSP48 & Register & LUT  \\
\hline
PPCF23 & 64.3 & 42.4 & 21.5\% & 100.0\% & 96.7\% & 828.3\%  \\
QAT+Pruning & 65.6 & 42.9 & 17.9\% & 100.0\% & 84.0\% & 399.9\%  \\
Optimized & 6.6 & 5.2 & 9.3\% & 7.8\% & 10.9\% & 36.7\%  \\
\bottomrule
\end{tabular}
\end{ruledtabular}
\caption{Comparison of latency\yw{, initiation interval} and resource usage estimates from logic syntheses for the three candidate models.  Resource usage estimates are shown as percentage of the respective type of resource available on the KU035 FPGA. 
}
\label{tab:finalmodel}
\end{table*}


\section{Hardware implementation}\label{sec:implementation}

\subsection{Firmware integration}

\begin{figure}
\includegraphics[width=0.99\linewidth]{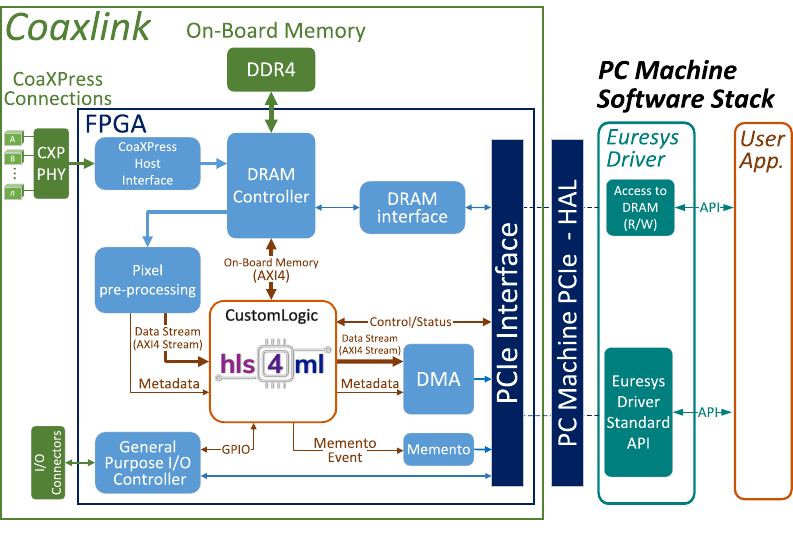}%
\caption{Euresys CustomLogic framework block diagram. \yw{Adapted with permission from Euresys, D209ET-Coaxlink CustomLogic User Guide-eGrabber 16.0.2.2128, Euresys S.A., Seraing, Belgium (2021).}}
\label{fig:customlogic}
\end{figure}

After compiling the optimized model to HLS representation, we integrated the neural network IP (intellectual property) block into the larger system firmware infrastructure and readout chain. 
Figure~\ref{fig:customlogic} shows the firmware block diagram of the Euresys Coaxlink Octo frame grabber.  Image data transmitted over the CoaXPress interface enters the diagram through the CoaXPress CXP-6 connectors (green, upper left).  The pixel data is then buffered before moving to the pixel pre-processing block and then streamed to the CustomLogic block where we integrate our synthesized \texttt{hls4ml} neural network block.  The output (and input image) can then be retrieved on the host computer over PCIe and over the general purpose I/O (GPIO) pins.

To integrate the neural network into the reference design, we leverage HLS to create the appropriate top level ports using properly sized input data types, implemented with an AXI-stream interface~\cite{axi_stream}. To ensure we extract and provide the correct data to the model, we track the current frame line and pixel position relative to the data stream using positional metadata provided with each data packet. Our current design duplicates the data stream, forwarding the original image to the PCIe interface and the duplicate to the model input. As a result, model inference continues in parallel with uninterrupted frame acquisition. Each data packet comprising the image contains eight pixels which divides the total time to input the image to the network by a factor of eight. 

\begin{figure}[b]
\includegraphics[width=0.99\linewidth]{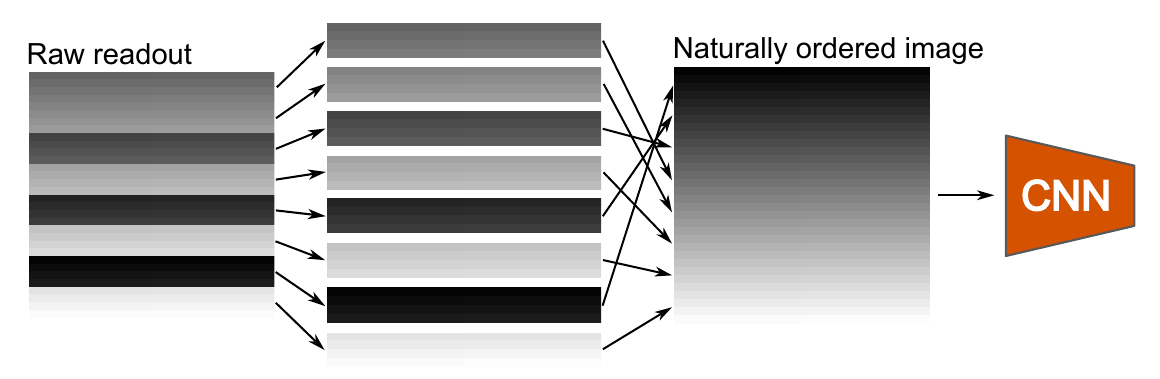}%
\caption{Pixel stream reordering operation implemented in the FPGA firmware.}
\label{fig:systemchallenges}
\end{figure}

We note a number of design challenges encountered when integrating our model.
Generally, meeting the camera protocol's timing constraint of 250~MHz was a challenge which we overcame \sout{with} \rff{through use of the} physical optimization tools \rff{in the Vivado Design Suite, as well as} \sout{and} hyperparameter searches of the best \rff{Vivado} synthesis strategies. 
Additionally, the ordering of the image lines within the data stream from the CoaXPress protocol IP do not match the natural (top-left to bottom-right) order of the captured image, but is instead divided into “stripes” (see Figure~\ref{fig:systemchallenges}). Since our neural network expects a naturally ordered input, and since first-in first-out (FIFO) memories are not addressable, we implement a stream reordering operation using intermediate FIFOs to store the split input stream before they are emptied in natural order. 
\yw{Furthermore, the Phantom S710 camera has a built-in restriction on the minimum width of the captured image which is 128 pixels. For this reason we decided to set the camera to capture at $128\times32$ pixels resolution, then in the firmware we track and pass only data packets belonging to the center $32\times32$ region of interest to the model input.}


\paragraph*{Control system outputs:}

We then implement the control request output of the frame grabber with an ap\_vld \sout{interface} \rff{port-level protocol. This protocol asserts a separate "valid" signal to alert the downstream endpoint when a new valid output is produced.} These control requests are calculated in the firmware through a single matrix multiplication using:

\begin{equation}
    v_{i,t} = g[s_t sin(m\theta_i + n\phi_i + \gamma) + c_t cos(m\theta_i + n\phi_i + \gamma)]
\end{equation}

\yw{\noindent where $v_{i,t}$ denotes the request for the applied field of control coil $i$ at time $t$,} $s_t$, $c_t$ are the sine and cosine components of the $n$=1 mode predicted by the neural network model \yw{at time $t$}, $\theta_i$ and $\phi_i$ are the poloidal and toroidal locations of control coil \yw{$i$}, $m$, and $n$ are the mode shape we intend to control, and $g$ and $\gamma$ are the gain and phase shift parameters to be scanned during a feedback control run campaign \cite{Peng16PPCF}. 
In the current implementation, the gain and phase shift factors are set to fixed values; for future versions of the firmware we would need the ability to adjust these values quickly between two consecutive discharges\ywsecond{.}

The control coils on HBT-EP are grouped into four toroidal arrays, each consisting of ten coils spaced evenly in the toroidal direction \yw{\cite{Maurer11PPCF}}. These four arrays are independent, but can also be hardwired to create phase shifts to generate arbitrary poloidal mode shape between $m$=1--4. Since we are targeting the $n$=1 mode, we can significantly simplify the problem by outputting only 5 control requests for half of one toroidal array (5 neighboring coils in one array) and mapping these five signals to all 40 coils using a breakout board to generate the desired mode shape. Note that this setup would only work when applying feedback control to a single $m/n$ mode. Multi-mode control tasks would require either switching the mode shape in the middle of a shot, or generating the requests through alternative methods. These will be explored in future studies on HBT-EP.

After computing the control requests, we map their decimal
representations to a 12-bit unsigned range\rffsecond{, append four DAC
control bits,} and transmit each \jplsecond{of the 5 requests serially over a pair\soutsecond{request serially over 5 pairs}} of RS422 differential output ports on the frame grabber board which are capable of speeds up to 10 MHz. The frame grabber will also supply the necessary serial clock and chip select control signals which are required for the DAC chip we will use for the future system. An additional on/off signal is mapped to an output port to indicate the beginning and conclusion of model inference. This allows us to empirically measure inference latency using an oscilloscope.

\paragraph*{Validation and verification:} 

For this work, we also developed a C testbench which logs the model output, input cropping and reordering, control request generation, and unsigned request mapping, allowing us to rapidly test and verify the HLS design. However, while this tool enables verification of the algorithms implemented in HLS prior to synthesis, we must also validate the results in hardware.

\begin{figure}
\centering
\includegraphics[width=1\linewidth]{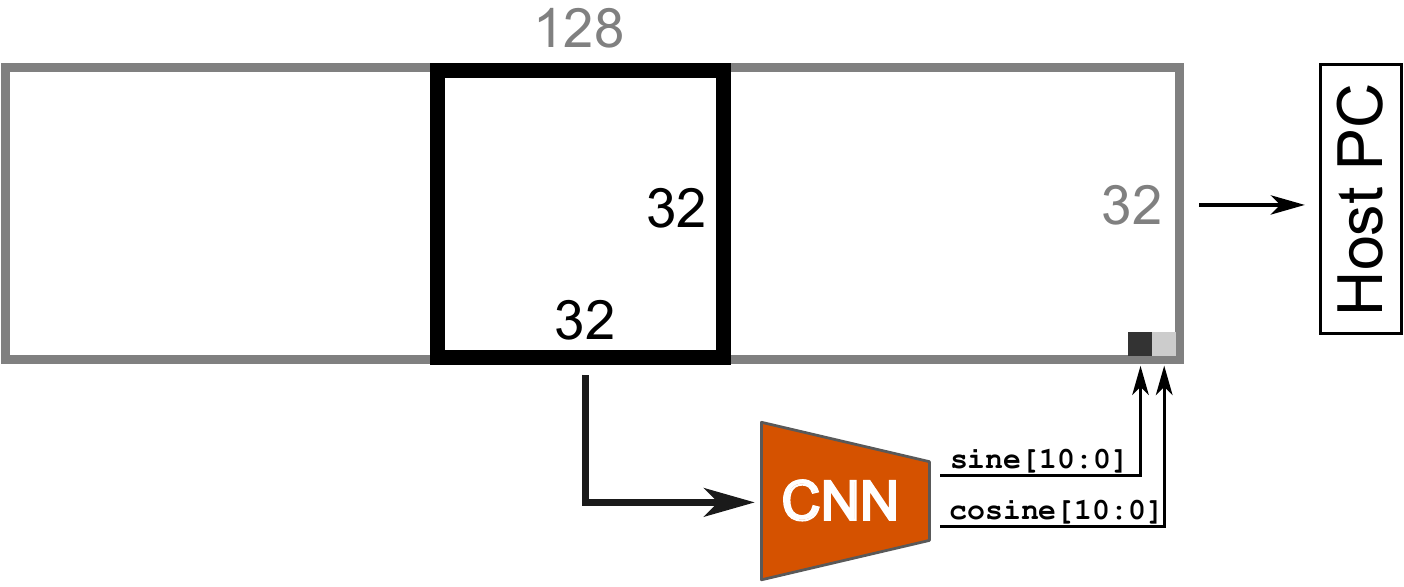}%
\caption{Method for model validation. The outputs of the firmware model are embedded to the lower right region of the captured $128\times32$ pixels resolution image, outside the actual $32\times32$ pixels region used by the implemented CNN model. This image is subsequently transferred over PCIe to the host-PC where the embedded predictions are compared with that of the same model running in a Python environment.}%
\label{fig:modelvalidation}
\end{figure}

Translating and verifying model results with its software counterpart using a bitwise writeout is time-consuming. Instead, we developed a significantly more scalable solution to validate the accuracy of our models across seconds’ worth of image acquisition. As our model considers only the center \yw{$32\times32$} crop of the full \yw{$128\times32$} acquisition, we have designed a variant of our firmware that embeds the model predictions into the outer region of the image (see Figure~\ref{fig:modelvalidation}). These modified images which carry their corresponding model predictions are then transferred over PCIe to the host computer’s memory. This approach enables us to validate the firmware accuracy over large amounts of data using a simple Python script. This output embedding is removed in the final firmware version to reduce latency and resource usage.  \sout{The implementation does not require modifications to the eGrabber API to retrieve the model predictions, which is left for future efforts.}


\subsection{System-level results}

\begin{figure}
\includegraphics[clip,width=0.99\columnwidth]{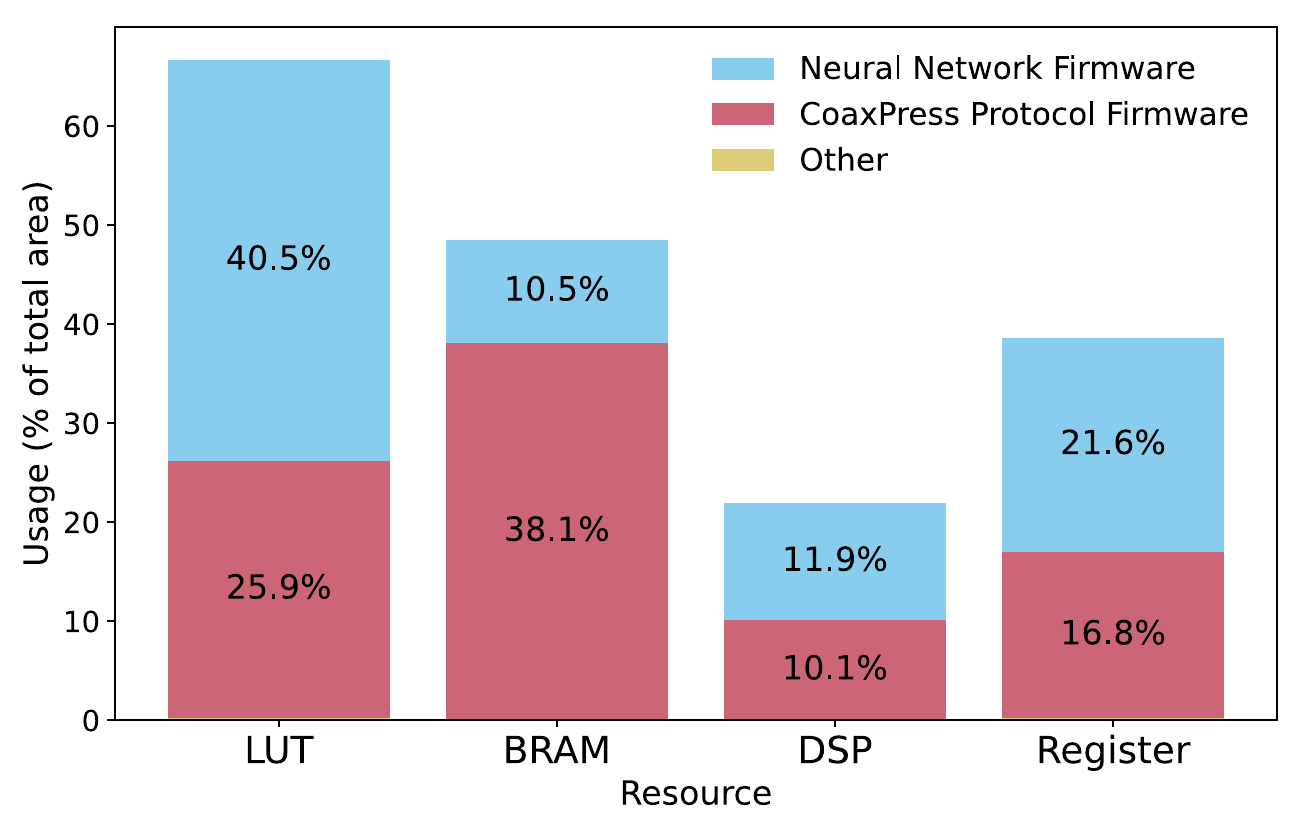}%
\\
\includegraphics[clip,width=0.99\columnwidth]{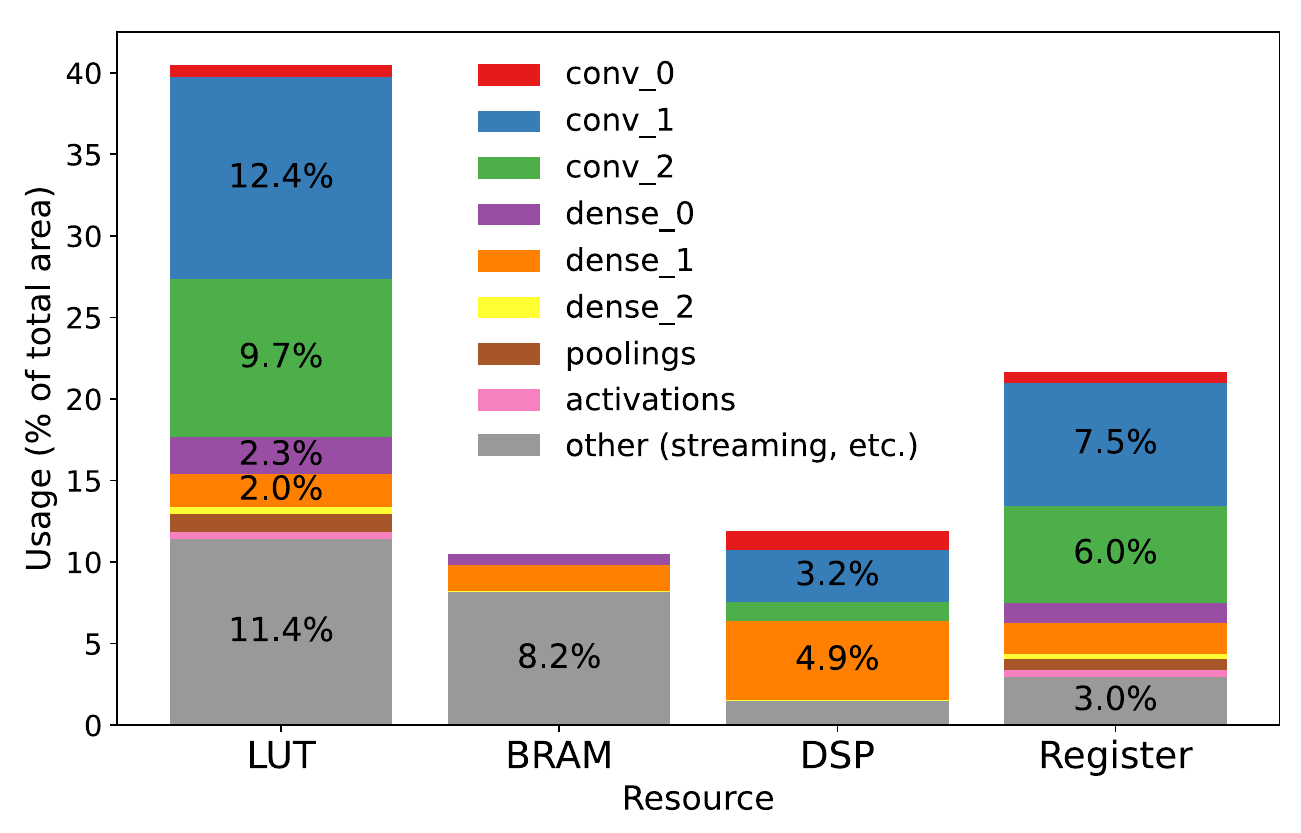}%
\caption{Frame grabber FPGA resource breakdown -- Top: overall IP core usage and breakdown between CoaxPress protocol and neural network.  Bottom: Resource usage breakdown within the neural network IP across the model architecture.}
\label{fig:resources}
\end{figure}

\begin{figure*}[htp]
\includegraphics[clip,width=0.9\linewidth]{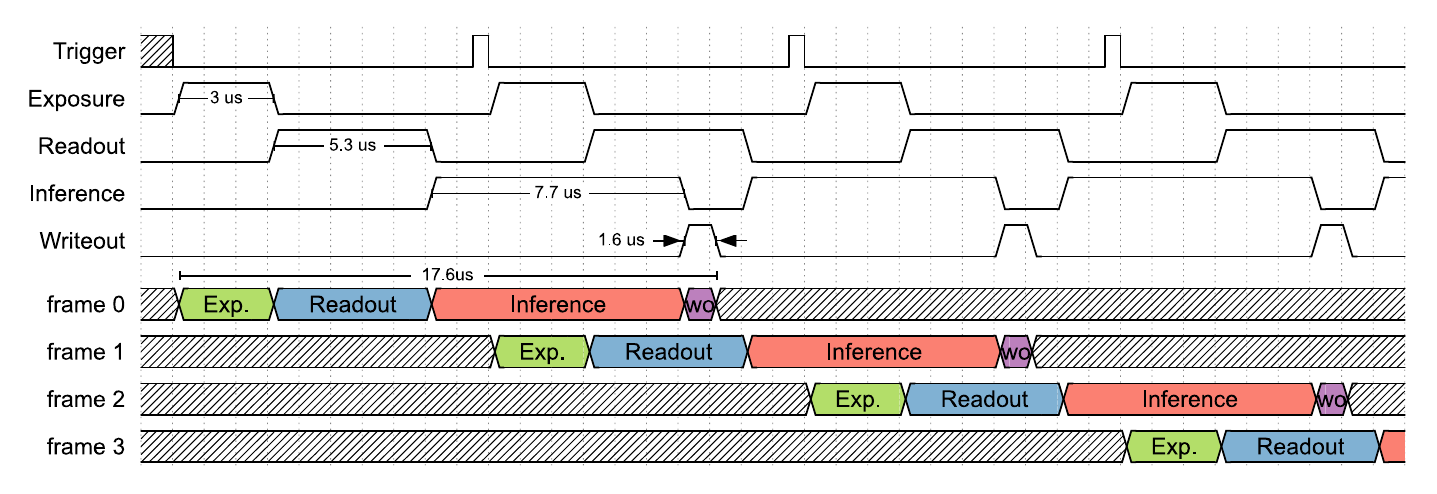} \\
\includegraphics[clip,width=0.9\linewidth]{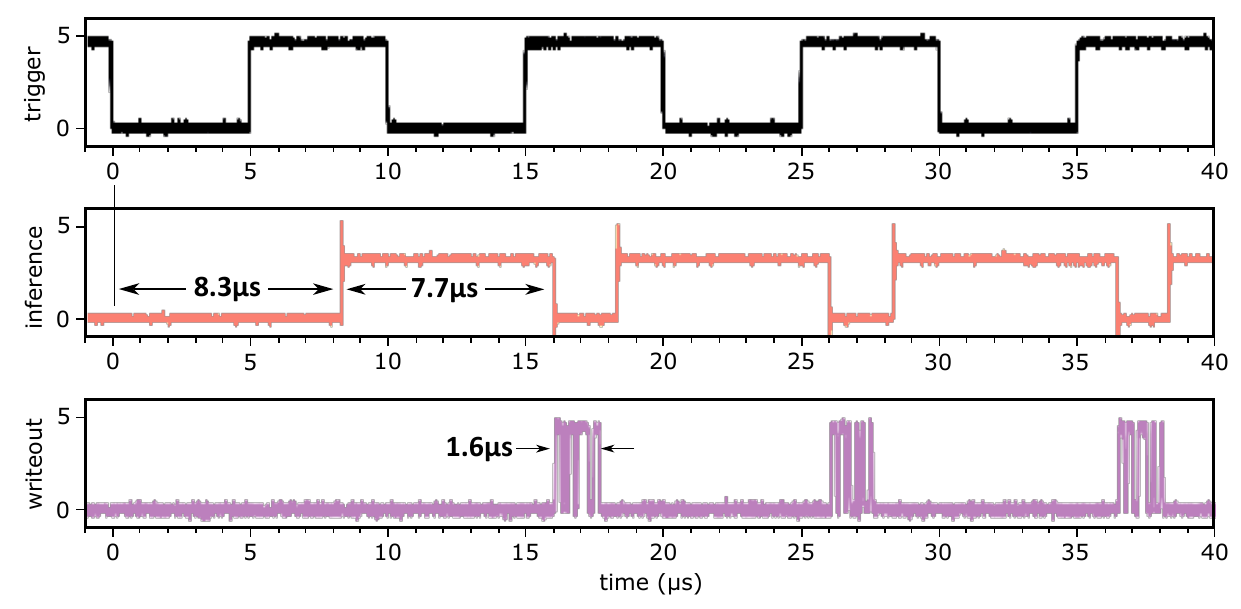}%
\caption{Frame grabber process timing diagram -- Top: Acquisition timing diagram at 100~kfps illustrating pipelined image capture and inference. ``Exp.'' denotes camera exposure, and ``wo'' denotes serial writout operation. Bottom: Oscilloscope latency measurement on the prototype platform at 100~kfps validating the timing diagram.}
\label{fig:waves}
\end{figure*}


The final resource consumption breakdown for the final implementation on the frame grabber FPGA (post-placement and routing) is shown in Figure~\ref{fig:resources}.  The resource usages are divided into three categories: the neural network model (includes cropping logic and control request calculation), the CoaXPress protocol logic, and other miscellaneous components. The neural network resource consumption is broken down further by layer and an additional ``other'' category which accounts for components such as BRAM-based FIFO data streams (these constitute the model input and the channels to transport activations from one layer to the next). Among the neural network layers, the convolutional layers require the most logic separate from multiply-accumulates, and therefore use the highest percentage of LUTs. The dense layers possess the highest parameter count, and therefore use the highest percentage of BRAMs for weight storage.

We deployed the final firmware on the frame grabber card and measured the total latency of the system on an oscilloscope.
Figure~\ref{fig:waves} (bottom) shows the oscilloscope measurements over a period of four frames at 100 kfps beginning with a camera capture trigger and ending with the final control output; an equivalent timing diagram is shown in Figure~\ref{fig:waves} (top). This system achieves a trigger-to-output latency of 17.6~\mbox{$\mu$s}, in which the 3~\mbox{$\mu$s} are attributed to camera exposure, 5.3~\mbox{$\mu$s} to camera readout and \yw{dynamic random-access memory} buffering, 7.7~\mbox{$\mu$s} to model inference, and finally 1.6~\mbox{$\mu$s} to serial writeout operation. Through pipelining model inference with the exposure and readout processes, we achieve throughputs of up to 120~kfps which corresponds to a minimum sampling interval of 8.3~$\mu$s. These results satisfy the latency and sampling interval requirements for performing real-time feedback control on HBT-EP and are only slightly slower compared to the previous GPU-based system \yw{(latency: 16~\mbox{$\mu$s}, sampling intervals: 4--6~\mbox{$\mu$s})} \cite{Rath13PPCF, Peng16PPCF}.






\section{Discussion and future work}\label{sec:discussion}


In this study, we have demonstrated an end-to-end workflow for implementing a deep learning model on an FPGA device for tracking the $n$=1 MHD mode evolution using high-speed camera data.
Our system achieves a total latency of 17.6~\mbox{$\mu$s} in which 7.7~\mbox{$\mu$s} is attributed to CNN model inference. Through pipelined operation, our system is capable to operate at up to 120~kfps. These satisfy the requirements for performing real-time feedback control of the $n$=1 MHD mode on HBT-EP. This is also the first demonstration of high-throughput microsecond-level low-latency implementation of a deep learning model for real-time diagnostic and control for plasma and fusion application.


The demonstrated workflow involves optimizing the CNN model in consideration with the available resources in the KU035 FPGA onboard the Euresys Coaxlink Octo frame grabber card. We used the open-source \texttt{hls4ml} package and optimization techniques including hyperparemeter optimization, quantization, pruning, and parallelization through setting reuse factors. This co-design process allowed us to meet both the system latency and FPGA resource requirements. 


Final integration and demonstration of the camera-and-FPGA-based real-time feedback control system requires addressing a few remaining points. 
To close the control loop requires an additional but simple DAC and breakout board system (Figure~\ref{fig:overview}) to interface the camera and frame grabber card implemented in this study with the downstream control coil actuators on HBT-EP. This additional system will take in the 5 control requests generated by the frame grabber card and map them to 40 coil voltage signals to generate a pre-selected mode shape. \jplsecond{A prototype DAC has been built for one of the control requests. \soutsecond{This is currently being designed and will be tested in the near future.}}
In addition, recent machine and diagnostic upgrades have changed the boundary conditions of HBT-EP plasma and thus making the current data set and model obsolete for the current plasma. For this reason, additional training data needs to be collected for updating the neural network parameters in the FPGA firmware prior to the control experiments. This, however, should not affect the performance of the FPGA-based mode tracking system in terms of latency, throughput, and system design and resources.

The last point on model retraining and adaptability to changing conditions is of particular interest for future studies.  The attraction of machine learning is to be able to adapt the algorithm for system needs and continuous running.  The current algorithm is effectively performing a regression task to extract the MHD mode signals of the plasma.  However, there are a number of novel directions for real-time control such as extending the platform for continuous monitoring and active learning, including domain adaptation or robustness methods in the model training, or changing the paradigm from supervised learning, which requires truth information of magnetic sensors, to real-time reinforcement learning which learns directly from the running system or in a simulator environment.  Our study opens these possibilities at the fastest speeds and the natural timescales of the most challenging plasma dynamics.  


Finally, the real-time camera-and-deep-learning-based diagnostic system demonstrated in this study\rff{\sout{could have} has} many applications beyond plasma and fusion research. \rff{Applications currently in progress include high-speed cancer cell classification and sorting\cite{NITTA2018266}, high-energy electron diffraction Gaussian fitting \cite{agar_applications}, and generic object detection and classification.} This system takes off-the-shelf camera and frame grabber components, uses standard data streaming protocols, and applies open-source packages and workflows for optimizing and deploying a neural network model \textit{in situ}. We believe this platform brings new scientific instrumentation capabilities to a wide range of application domains and is accessible to a wide audience.


\section*{Acknowledgments}\label{sec:acknowledgments}

We acknowledge the Fast Machine Learning collective as an open community of multi-domain experts and collaborators. This community and Jovan Mitrevski, Jonathan Eisch, and Christian Herwig, in particular were important for the development of this project.

Y. Wei is supported by U.S. DOE, Office of Fusion Energy Science, Grant DE-FG02-86ER53222 and DE-SC0022234.
C. Hansen is supported by U.S. DOE, Office of Fusion Energy Science, Grant DE-SC0021325.
J. P. Levesque, M. E. Mauel, and G. A. Navratil are supported by U.S. DOE, Office of Fusion Energy Science, DE-FG02-86ER53222.
N. Tran is supported by Fermi Research Alliance, LLC under Contract No. DE-AC02-07CH11359 with the Department of Energy (DOE), Office of Science, Office of High Energy Physics.  
R. F. Forelli, N. Tran, and J. C. Agar are also supported by the DOE Office of Science, Office of Advanced Scientific Computing Research under the “Real-time Data Reduction Codesign at the Extreme Edge for Science” Project (DE-FOA-0002501).
Computational resources were in part provided by NSF MRI, 2320600 MRI: Track 2 Development of a Platform for Accessible Data-Intensive Science and Engineering and 2215789 MRI: Development of Heterogeneous Edge Computing Platform for Real-Time Scientific Machine Learning.

\section*{Data Availability}\label{sec:dataavailability}
The data that support the findings of this study are available upon reasonable request from the authors.

\section*{References}\label{sec:references}



%
%

%


\bibliography{fastcam-hls4ml-paper_current}

\end{document}